\documentclass[doublespacing]{elsart}

\usepackage{latexsym}
\usepackage{amsmath}
\usepackage{amssymb}
\usepackage{amscd}
\usepackage{bm}
\usepackage{graphicx}

\begin{document}

\begin{frontmatter}



\title{Greenberger-Horne-Zeilinger correlation and Bell-type inequality seen from moving frame}


\author[label1]{Hao You\thanksref{email}}
\thanks[email]{E-mail:youhaoca@mail.ustc.edu.cn}
\author[label1,label2]{An Min Wang}
\author[label1]{Xiaodong Yang}
\author[label1]{Wanqing Niu}
\author[label1]{Xiaosan Ma}
\author[label1]{Feng Xu}
\address[label1]{Department of Modern Physics, University of
Science and Technology of China, Hefei 230026, People's Republic
of China}
\address[label2]{Laboratory of Quantum Communication and Quantum Computing and Institute for
 Theoretical Physics, University of Science and Technology of China, Hefei 230026, People's Republic of China}

\begin{abstract}
The relativistic version of the Greenberger-Horne-Zeilinger
experiment with massive particles is proposed. We point out that,
in the moving frame, GHZ correlations of spins in original
directions transfer to different directions due to the Wigner
rotation. Its effect on the degree of violation of Bell-type
inequality is also discussed.
 \end{abstract}

\begin{keyword}
GHZ correlation \sep Bell-type inequality \sep relativity
\PACS 03.65.Bz \sep 03.30.+p
\end{keyword}
\end{frontmatter}


Contemporary applications of the Einstein, Podolsky, and Rosen
(EPR) correlations and the Bell inequality range from purely
theoretical problems \cite{EPR,Bohm,Bell} to quantum communication
such as quantum teleportation \cite{Bennett1} and quantum
cryptography \cite{Ekert,Bennett2}. Recently a lot of interest has
been devoted to the study of the EPR correlation function under
the Lorentz transformations
\cite{Czachor,Terashima,Terashima1,Ahn1,Ahn2,Terashima2}. They
showed that, the relativistic effects on the EPR correlations are
nontrivial and the degree of violation of the Bell inequality
depends on the relative motion of the particles and the observers.

Greenberger, Horne and Zeilinger (GHZ) also proposed a kind of
quantum correlations known as GHZ correlations
\cite{Greenberger1,Bouwmeester1}. This kind of refutation of local
realism is strikingly more powerful than the one Bell's theorem
provides for Bohm's version of EPR --- it is no longer statistical
in principle. Furthermore, GHZ correlations are essential for most
quantum communication schemes in practice
\cite{Bouwmeester1,pan,Bouwmeester2,Nelson}. Thus it is an
interesting question that whether GHZ correlations can still be
held in the moving frame.

In this letter, we formulate a relativistic GHZ
gedanken-experiment with massive particles considering a situation
in which measurements are performed by moving observers. It is
pointed that GHZ correlations of spins in original directions no
longer hold and transfer to different directions in the moving
frame. This is a consequence of Wigner rotation \cite{Wigner} and
does not imply a breakdown of non-local correlation. To obtain and
utilize perfect properties of GHZ correlations again, we should
choose spin variables to be measured appropriately to purse
desired tasks. Our intention is to explore effects of the relative
motion between the sender and receiver which may play role in
future relativistic experiment testing the strong conflict between
local realism and quantum mechanics, or which may be useful in
future quantum information processing using GHZ correlations in
high velocity case.

The version of the GHZ experiment in non-relativistic case is
listed as follow \cite{Bouwmeester1,Mermin1}. Consider three
spin-$\frac{1}{2}$ particles prepared in the state
\begin{equation}
|\psi\rangle=\frac{1}{\sqrt{2}}[|\uparrow;\uparrow;\uparrow\rangle+
|\downarrow;\downarrow;\downarrow\rangle]
\end{equation}
Here $|\uparrow\rangle$ represents "up" along the $z$ axis and
$|\downarrow\rangle$ signifies "down" along the $z$ axis. Now
consider the result of the following products of spin
measurements, each made on state $|\psi\rangle$.\\ (i) Particle
$1$ along $y$, particle $2$ along $y$, particle $3$ along $x$.
Note that since
$\sigma_{y}\sigma_{y}\sigma_{x}|\psi\rangle=-|\psi\rangle$. The
product of the $yyx$ measurements should be $-1$, i.e., the
expectation value of three-particle spin correlation in the
direction $yyx$
(which is denoted by $E(yyx)$), should be $-1$.\\
(ii) Particle $1$ along $y$, particle $2$ along $x$, particle $3$
along $y$. The product of the $yxy$ measurements $E(yxy)$ should be $-1$.\\
(iii) Particle $1$ along $x$, particle $2$ along $y$, particle $3$
along $y$. The product of the $xyy$ measurements $E(xyy)$ should be $-1$.\\
(iv) Particle $1$ along $x$, particle $2$ along $x$, particle $3$
along $x$. In this case, the product of the $xxx$ measurements
$E(xxx)$ should be $+1$.

Obviously spin correlations in directions $yyx$, $yxy$, $xyy$, and
$xxx$ are maximally correlated, known as GHZ correlations. And the
positive sign in the final scenario is crucial for differentiating
between quantum-mechanics and hidden-variable descriptions of
reality, because local realistic theory predicts the product be
$-1$. We can see whenever local realism predicts that a specific
result definitely occurs for a measurement on one of the
particle's spin given the results for the other two, quantum
physics definitely predicts the opposite result. Thus, using GHZ
correlations, quantum mechanics predictions are in conflict with
local realism definitely, while in the case of EPR experiments,
quantum mechanics predictions are in conflict with local realism
only statistically. In experiment, GHZ's prediction was confirmed
in Ref.\cite{Nelson} using nuclear magnetic resonance.

In relativistic case, the Lorentz transformation induces unitary
transformation on vectors in Hilbert space\cite{Wigner}. Suppose
that a massive spin-$\frac{1}{2}$ particle moves with the
laboratory-frame $4$-momentum
$p=(m\cosh\xi,m\sinh\xi\sin\theta\cos\phi,\\
m\sinh\xi\sin\theta\sin\phi,m\sinh\xi\cos\theta)$ where the
rapidity $\vec{\xi}=\xi\hat{\textbf{p}}$ with the normal vector
$\hat{\bf{p}}=(\sin\theta\cos\phi,\sin\theta\sin\phi,\cos\theta)$.
An observer is moving along $z$-axis with the velocity $\vec{V}$
in the laboratory frame. The rest frame of the observer is
obtained by performing a Lorentz transformation
$\Lambda=L^{-1}(\vec{\chi})=L(-\vec{\chi})$ on the laboratory
frame with the rapidity $-\vec{\chi}=\chi(-\hat{\textbf{z}})$.
Here $V=\tanh\chi$ and $-\hat{\textbf{z}}=(0,0,-1)$ is the normal
vector in the boost direction. In this frame, the observer
describes the $4$-momentum eigenstate $|p\lambda\rangle$ as
$U(\Lambda)|p\lambda\rangle$ (for a review of momentum
eigenstates, one may refer to
Ref.\cite{Weinberg,Peres1,Gingrich}). A straightforward
calculation shows that \cite{Terashima,Terashima1,Ahn1,Ahn2}:
\begin{eqnarray}
U(\Lambda)|p\uparrow\rangle&=&
\cos\frac{\delta}{2}|\Lambda p\uparrow\rangle+e^{i\phi}\sin\frac{\delta}{2}|\Lambda p\downarrow\rangle\\
U(\Lambda)|p\downarrow\rangle&=&
-e^{-i\phi}\sin\frac{\delta}{2}|\Lambda
p\uparrow\rangle+\cos\frac{\delta}{2}|\Lambda p\downarrow\rangle
\end{eqnarray}
where $\uparrow$ and $\downarrow$ represent "up" and "down" along
$z$-axis, respectively. The Wigner rotation is indeed a rotation
about the direction $\hat{\delta}=
(-\bf{\hat{\textbf{z}}}\times\hat{\textbf{p}})/|\bf{\hat{\textbf{z}}}\times\hat{\textbf{p}}|$
through the angle $\delta$:
\begin{eqnarray}
\cos\delta&=&\frac{A-B(\hat{\textbf{z}}\cdot\hat{\textbf{p}})+C(\hat{\textbf{z}}\cdot\hat{\textbf{p}})^{2}}
{D-B(\hat{\textbf{z}}\cdot\hat{\textbf{p}})}\\
\sin\delta\hat{\delta}&=&-\frac{B-C(\hat{\textbf{z}}\cdot\hat{\textbf{p}})}{D-B(\hat{\textbf{z}}\cdot\hat{\textbf{p}})}\hat{\textbf{z}}\times\hat{\textbf{p}}
\end{eqnarray}
with
\begin{eqnarray}
A&=&\cosh\xi+\cosh\chi\\
B&=&\sinh\xi\sinh\chi\\
C&=&(\cosh\xi-1)(\cosh\chi-1)\\
D&=&\cosh\xi\cosh\chi+1
\end{eqnarray}

A GHZ state for three massive particles in the laboratory frame
reads:
\begin{equation}
|\psi\rangle=\frac{1}{\sqrt{2}}[|p_{1}\uparrow;p_{2}\uparrow;p_{3}\uparrow\rangle+
|p_{1}\downarrow;p_{2}\downarrow;p_{3}\downarrow\rangle]
\end{equation}
where
$p_{i}=(m\cosh\xi_{i},m\sinh\xi_{i}\sin\theta_{i}\cos\phi_{i},
m\sinh\xi_{i}\sin\theta_{i}\sin\phi_{i},m\sinh\xi_{i}\cos\theta_{i})$
represents the $4$-momentum of the $i$-th particle in the
laboratory frame and $i=1,2,3$. Without loss of generality,
$\phi_{3}$ is set to $0$ which means the third particle moves in
the $yOz$-plane. It is necessary to explicitly specify the motion
of the particles because Wigner rotation depends on the momentum.
In this GHZ experiment, suppose three spin-$\frac{1}{2}$
particles, prepared in the state (10), move apart from the GHZ
source and are detected by three observers. Each observer measures
a spin component along a chosen direction. Note that whatever
frame is chosen for defining simultaneity, the experimentally
observable result is the same \cite{Peres1,Peres2}, so we needn't
discuss the chronology of spin measurements. Here we assume that,
three observers are moving in the $z$ direction at the same
velocity $\vec{V}$ in the laboratory frame. What we are interested
in are GHZ correlations in the common inertial frame where the
observers are all at rest. In this moving frame, the observers see
the GHZ state (10) as:
\begin{multline}
U(\Lambda)|\psi\rangle=
\frac{1}{\sqrt{2}}\bigg[\big(c_{1}c_{2}c_{3}-e^{-i(\phi_{1}+\phi_{2})}s_{1}s_{2}s_{3}\big)
|\Lambda p_{1}\uparrow;\Lambda p_{2}\uparrow;\Lambda p_{3},\uparrow\rangle\\
+\big(e^{-i(\phi_{1}+\phi_{2})}s_{1}s_{2}c_{3}+c_{1}c_{2}s_{3}\big)
|\Lambda p_{1}\uparrow;\Lambda p_{2}\uparrow;\Lambda
p_{3}\downarrow\rangle\\
+\big(e^{-i\phi_{1}}s_{1}c_{2}s_{3}+e^{i\phi_{2}}c_{1}s_{2}c_{3}\big)
|\Lambda p_{1}\uparrow;\Lambda p_{2}\downarrow;\Lambda p_{3}\uparrow\rangle\\
+\big(-e^{-i\phi_{1}}s_{1}c_{2}c_{3}+e^{i\phi_{2}}c_{1}s_{2}s_{3}\big)
|\Lambda p_{1}\uparrow;\Lambda p_{2}\downarrow;\Lambda
p_{3}\downarrow\rangle\\
+\big(e^{-i\phi_{2}}c_{1}s_{2}s_{3}+e^{i\phi_{1}}s_{1}c_{2}c_{3}\big)
|\Lambda p_{1}\downarrow;\Lambda p_{2}\uparrow;\Lambda p_{3}\uparrow\rangle\\
+\big(-e^{-i\phi_{2}}c_{1}s_{2}c_{3}+e^{i\phi_{1}}s_{1}c_{2}s_{3}\big)
|\Lambda p_{1}\downarrow;\Lambda p_{2}\uparrow;\Lambda
p_{3}\downarrow\rangle\\
+\big(e^{i(\phi_{1}+\phi_{2})}s_{1}s_{2}c_{3}-c_{1}c_{2}s_{3}\big)
|\Lambda p_{1}\downarrow;\Lambda p_{2}\downarrow;\Lambda p_{3}\uparrow\rangle\\
+\big(c_{1}c_{2}c_{3}+e^{i(\phi_{1}+\phi_{2})}s_{1}s_{2}s_{3}\big)
|\Lambda p_{1}\downarrow;\Lambda
p_{2}\downarrow;\Lambda p_{3}\downarrow\rangle\Big]\\
\end{multline}
where
$c_{i}\equiv\cos{\frac{\delta_{i}}{2}},s_{i}\equiv\sin{\frac{\delta_{i}}{2}}.$
 And $\delta_{i}$ represents the Wigner angle of the \emph{i}-th particle
which is defined in (4) and (5). Spin operators in relativistic
case are defined as \cite{Terashima,Terashima1,Terashima2}:
\begin{eqnarray}
\sigma_{x}(p)&=&|p\uparrow\rangle\langle p\downarrow|+|p\downarrow\rangle\langle p\uparrow|\\
\sigma_{y}(p)&=&-i|p\uparrow\rangle\langle p\downarrow|+i|p\downarrow\rangle\langle p \uparrow|\\
\sigma_{z}(p)&=&|p\uparrow\rangle\langle
p\uparrow|-|p\downarrow\rangle\langle p\downarrow|
\end{eqnarray}
Since the observers are moving in the $z$ direction, the
directions that are parallel in the laboratory frame remain
parallel in the moving frame where the observers are all at rest.
However, whether the results of spin measurements in the same
direction are still maximally correlated in this moving frame
isn't obvious. We now research this question. Let the observer who
receives particle $1$ performs measurement of $\sigma_{y}$, the
observer who receives particle $2$ performs measurement of
$\sigma_{y}$, and the observer who receives particle $3$ performs
measurement of $\sigma_{x}$. Thus in the moving frame, the
expectation value of three-particle spin correlation in the
direction $yyx$ is obtained:
\begin{equation}
E(yyx)=-\cos\delta_{3}
\end{equation}
Similarly we obtain:
\begin{eqnarray}
E(yxy)&=&-\cos\delta_{2}\\
E(xyy)&=&-\cos\delta_{1}
\end{eqnarray}
We can see that GHZ correlations that are maximally correlated in
the laboratory frame no longer appear so in the moving frame. That
is, in the moving frame, given the results of measurements on two
particles, one can't predict with certainty what the result of a
corresponding measurement performed on the third particle. In
practice, this means that the relative motion between the source
of entangled particles and the observers can alter properties of
spin correlations when the observers receive the particles. Thus
quantum information processing using these perfect correlations of
the GHZ state can't be held, for example, the GHZ experiment can't
be done due to lack of knowledge of GHZ correlations in the moving
frame.

This effect occurs because Lorentz transformation rotates the
direction of spin of the particle as can be seen form (2) and (3).
Since the Wigner rotation is in fact a kind of local
transformation, it preserves the entanglement of the state
\cite{Alsing}. Thus it is reasonable that the GHZ correlation
should be preserved in appropriately chosen direction. Here we
point out that, to utilize GHZ correlations in the moving frame,
the observers should choose spin variables to be measured
appropriately according to the wigner rotation:
\begin{eqnarray}
\sigma_{x}(\Lambda p_{i})\rightarrow \sigma_{x^{\prime}}(\Lambda
p_{i})&=&U(\Lambda)\sigma_{x}(\Lambda
p_{i})U(\Lambda)^{+}\nonumber\\&=&(c_{i}^{2}-s_{i}^{2}\cos{2\phi_{i}})\sigma_{x}(\Lambda
p_{i})-s_{i}^{2}\sin{2\phi_{i}}\sigma_{y}(\Lambda
p_{i})\nonumber\\& &-2s_{i}c_{i}\cos{\phi_{i}}\sigma_{z}(\Lambda
p_{i})\\
\sigma_{y}(\Lambda p_{i})\rightarrow \sigma_{y^{\prime}}(\Lambda
p_{i})&=&U(\Lambda)\sigma_{y}(\Lambda
p_{i})U(\Lambda)^{+}\nonumber\\&=&-s_{i}^{2}\sin{2\phi_{i}}\sigma_{x}(\Lambda
p_{i})+(c_{i}^{2}+s_{i}^{2}\cos{2\phi_{i}})\sigma_{y}(\Lambda
p_{i})\nonumber\\& &-2s_{i}c_{i}\sin{\phi_{i}}\sigma_{z}(\Lambda
p_{i})
\end{eqnarray}
where the index $i$ represents the $i$-th particle. Thus GHZ
correlations will be obtained in new different directions in the
moving frame. For example, if the observer who receives particle
$1$ measures spin along the direction
$y_{1}^{\prime}=(-s_{1}^{2}\sin{2\phi_{1}},c_{1}^{2}+s_{1}^{2}\cos{2\phi_{1}},-2s_{1}c_{1}\sin{\phi_{1}})$,
the observer who receives particle $2$ measures spin along the
direction
$y_{2}^{\prime}=(-s_{2}^{2}\sin{2\phi_{2}},c_{2}^{2}+s_{2}^{2}\cos{2\phi_{2}},-2s_{2}c_{2}\sin{\phi_{2}})$
and the observer who receives particle $3$ measures spin along the
direction $x_{3}^{\prime}=(c_{3}^{2}-s_{3}^{2},0,-2s_{3}c_{3})$,
maximal correlation $E(y^{\prime}y^{\prime}x^{\prime})=-1$ is
obtained again. That is, the GHZ correlation in the direction
$yyx$ in the laboratory frame transfers to new direction
$y^{\prime}y^{\prime}x^{\prime}$ seen from the moving frame.
Similar conclusions are also held for
$y^{\prime}x^{\prime}y^{\prime}$ and
$x^{\prime}y^{\prime}y^{\prime}$ cases with a careful choice of
spin variables according to (18) and (19).

Now we can perform GHZ experiment in the moving frame. After a set
of spin measurements along the direction
$y^{\prime}y^{\prime}x^{\prime}$,
$y^{\prime}x^{\prime}y^{\prime}$, and
$x^{\prime}y^{\prime}y^{\prime}$ respectively, local realism will
predict the possible outcomes for a
$x^{\prime}x^{\prime}x^{\prime}$ spin measurement must be those
terms yielding a expectation value
$E(x^{\prime}x^{\prime}x^{\prime})=-1$. While quantum theory
predicts the outcomes should be the terms yielding
$E(x^{\prime}x^{\prime}x^{\prime})=1$. Then the strong conflict
between the quantum theory and the local realism is seen in the
moving frame in principle.

Similarly can people test the Bell-type inequality for three-qubit
state in relativistic case. In non-relativistic case, any local
realistic theory predicts
$|\varepsilon|=|E(xyy)+E(yxy)+E(yyx)-E(xxx)|\leq2$ while the
maximal possible value is reached for the GHZ state where
$|\varepsilon|=4$ \cite{Mermin,Klyshko}. If the directions of
measurements of spin are fixed as $xyy$, $yxy$, $yyx$, and $xxx$,
the degree of violation for the GHZ state in the moving frame
where the observers are at rest equals to:
\begin{equation}
|\varepsilon|=4\sqrt{(c_{1}c_{2}c_{3})^4+(s_{1}s_{2}s_{3})^4-2(c_{1}c_{2}c_{3}s_{1}s_{2}s_{3})^2\cos{[2(\phi_{1}+\phi_{2})}]}
\end{equation}
The result depends on the velocity of both the particles and the
observers with respect to the laboratory in terms of parameters
$\theta$, $\phi$, and the Wigner angle $\delta$. If the particles
(or the observers) are all at rest in the laboratory frame or if
the moving direction of the observer is parallel with that of the
particle to be measured, the amount of violation reaches to the
maximal value $|\varepsilon|=4$ which gives the same outcome as
the case in non-relativistic case. It is interesting to see in
some cases the observers will find the degree of violation to be
zero. For example, In the case $\xi\rightarrow\infty$ and
$\chi\rightarrow\infty$ and $\theta=\pi/2$, where the particles
and the observers move perpendicularly with high velocities,
observers will find $|\varepsilon|=0$. In fact, if three particles
rotate angles that are represented by points on the surface
$\tan\frac{\delta_{1}}{2}\tan\frac{\delta_{2}}{2}\tan\frac{\delta_{3}}{2}=1$,
the degree of violation for the GHZ state is $0$ when
$\phi_{1}+\phi_{2}=n\pi$.

The change in the degree of violation of the Bell-type inequality
also results from the fact that the Wigner rotation rotates the
direction of spins and thus perfect correlations transfer to
different directions as we point out above. As can be seen from
(18) and (19), if observers rotate the directions of measurements
in accordance with the Wigner rotation, Bell-type inequality turns
out to be maximally violated with $|\varepsilon|=4$.

In summarize, We apply a specific Lorentz boost to the GHZ state
and then compute the expectation value of three-particle spin
correlations in the transformed state. As a result, spin variables
averages that are maximally correlated in the laboratory frame no
longer appear so in the same directions seen from the moving
frame. The entanglement of the GHZ state is however not lost and
perfect correlations of the GHZ state are always possible to be
found in different directions seen from the moving frame. As its
applications, we formulated GHZ experiment in relativistic case,
and Bell-type inequality for three-qubit in relativistic case is
also discussed. If the relative motion between the source of
entangled particles and the observers must be taken account of, we
should consider this GHZ correlation transfer in practice.

We thank Xiaoqiang Su, Ningbo Zhao, and Rengui Zhu for useful
discussions. We also thank the anonymous referees for pointing out
to us some obscure points and for their constructive comments.
This project was supported by the National Basic Research
Programme of China under Grant No 2001CB309310, the National
Natural Science Foundation of China under Grant No 60173047, the
Natural Science Foundation of Anhui Province, and China
Post-doctoral Science Foundation.

\end{document}